\documentclass[doublecol]{epl2} 

\title{Some new results on one-dimensional outflow dynamics.}

\author{F. Slanina\inst{1,2} \and K. Sznajd-Weron\inst{3} \and P. Przyby{\l}a \inst{3}}
\shortauthor{F. Slanina \etal}

\institute{                    
\inst{1} 
Institute of Physics,
Academy of Sciences of the Czech Republic,
Na~Slovance~2, CZ-18221~Praha,
Czech Republic\\
\inst{2}
Center for Theoretical Study, Jilsk\'a 1, Prague, Czech Republic\\
\inst{3} 
Institute of Theoretical Physics, University of
Wroc{\l}aw, pl. Maxa Borna 9, 50-204 Wroc{\l}aw, Poland
}


\abstract{
In this paper we introduce modified version of one-dimensional outflow
dynamics (known as a Sznajd model) which simplifies the analytical treatment. We show that
simulations results of the original and modified rules are exactly the
same for various initial conditions. We obtain the analytical formula
for exit probability using Kirkwood approximation and we show that it
agrees perfectly with computer simulations in case of random initial
conditions. Moreover, we compare our results with earlier analytical
calculations obtained from renormalization group and from general
sequential probabilistic frame introduced by Galam. Using computer
simulations we investigate the time evolution of several correlation
functions to show if Kirkwood approximation can be
justified. Surprisingly, it occurs that Kirkwood approximation gives
correct results even for these initial conditions for which it cannot
be easily justified.}  

\begin{document}

\maketitle

\section{Introduction}

The outflow dynamics was introduced to describe the opinion change in
the society. The idea was based on the fundamental social phenomenon
called "social validation". By now, the opinion dynamics was studied
by many authors, starting perhaps from the works by Galam
\cite{galam_90} and developed later in the Sznajd \cite{szn_szn_00}
and Majority rule \cite{kra_red_03} models. The common feature of
these models is that the complexity of real-world opinions is reduced
to the minimum set of two options, $+$ or $-$. 

However, in this paper we do not focus on
social applications of the model (an interested reader may resort to 
reviews  \cite{Stauffer02,Schechter02,FS05,SW05,CFL2007}). Here we
deal with a more mathematical problem, namely
finding the analytical formula for the probability $P_+(p)$ of
reaching consensus on opinion $+$ as a function of the initial
fraction $p$ of opinion $+$. This quantity is commonly 
called exit probability
\cite{mob_red_03,lr2007}. In fact, we follow the method used in
\cite{mob_red_03} for the Majority-rule model.

The 
one dimensional outflow dynamics is defined as follows: if pair of
neighboring spins $S_iS_{i+1}=1$ the the two neighbors of the pair
followed its direction, i.e. $S_{i-1} \rightarrow S_{i}(=S_{i+1})$ and
$S_{i+2} \rightarrow S_{i}(=S_{i+1})$; in case of different opinions
at the central pair, the two neighboring states are unchanged. Until
now several analytical approaches have been proposed. 
One of the analytical approaches used for the outflow dynamics was based on
the mean field idea \cite{SL03}. Within the mean field approach
  the mean relaxation time
$\langle\tau\rangle$ as a function of the initial fraction $p$ of
  opinion $+$ was 
  computed, as
well as distribution of relaxation times. The exit probability found
in this approach is the trivial step function, at odds with the known
simulation results for 1D dynamics. 

\begin{figure}
\onefigure[scale=0.4]{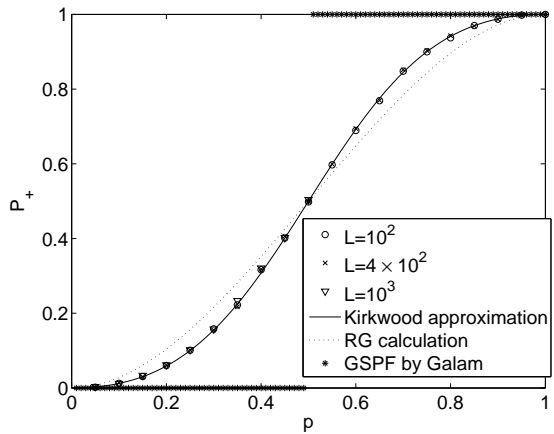}
\caption{Exit probability $P_+$ from random initial state consisting
  of $p$ up-spins for the modified original outflow dynamics in one
  dimension for several lattice sizes $L$. This is seen that results
  agree perfectly with analytical formula (solid line) given by
  eq.(\ref{exitp}) obtained from Kirkwood
  approximation. Renormalization group (RG) results obtained in
  \cite{GSH2006} for growing networks and calculations made by Galam
  within his general sequential probabilistic frame (GSPF) agree with
  simulation results much worse. Results obtained for modified version
  of outflow dynamics in which only one neighbor of central pair is
  changed are exactly the same. Results are averaged over $10^4$
  samples.} 
\label{fig:exitprob}
\end{figure}

Later, Galam in \cite{g05} presented a general sequential
probabilistic frame (GSPF), which extended a series of earlier opinion
dynamics models. Within his frame he was able to find  analytic
formulas for the probability $p(t + 1)$ to find at random an agent
sharing opinion $+$ at time $t+1$ as a function of $p(t)$. Among
several models, he considered the one dimensional rule, which we
investigate in this paper, i.e.: if pair of neighboring spins
$S_iS_{i+1}=1$ then the two neighbors of the pair followed its
direction; in case of different opinions at the central pair, the two
neighboring states are unchanged.  
For such a rule, within his GSPF calculation Galam has found the following formula \cite{g05}:
\begin{eqnarray}
p(t+1) & = & p(t)^4+\frac{7}{2}p(t)^3[1-p(t)] \nonumber\\
& + & 3p(t)^2[1-p(t)]^2+\frac{1}{2}p(t)[1-p(t)]^3
\end{eqnarray}
Iterating above formula the exit probability $P_+$ can be found as a
step function (see Fig. \ref{fig:exitprob}). It is worth to notice
that step-like function describes exit probability in the case of two
dimensional outflow dynamics \cite{SSO00}, but not in one dimension
\cite{SWK06}. Step-like function for exit probability has been found
also in \cite{GSH2006}. 

In the paper \cite{GSH2006} real space renormalization approach has
been proposed to calculate the probability $P_+(p)$ of reaching
consensus on opinion $+$ as a function of the initial 
fraction p of opinion $+$. They have found in case of two sites
convincing others the following analytical formulas: 
\begin{itemize}
\item
in the case of growing network (growing hierarchical or the Barabasi-Albert
scale-free network)
\begin{equation}
P_+=3p^2-2p^3.
\end{equation}
\item
in the case of fixed network they have found $P_+$ as the step
function observed also in computer simulations on the square lattice. 
\end{itemize}
It is seen on Fig. \ref{fig:exitprob} that RG results for growing
networks agree much better with simulations that RG results for fixed
network which are exactly the same as obtained using GSPF by Galam.  

In this paper we present analytical results obtained using Kirkwood
approximation \cite{mob_red_03} following the method used in  for the
Majority-rule model and we obtain perfect agreement with computer
simulations. Moreover, we consider two types of non-random initial
conditions and we show how analytical formulas for exit probabilities
change for such cases. 

\section{Approximate solution in 1D}

We consider linear chain with $L$ sites. Each site can be in two
states $\pm 1$. 
We use the following notation:\\
$\sigma\in \{-1,+1\}^L$ state of the system.\\
$\sigma(y)$ state of the spin at site $y$ if the system is in state $\sigma$.\\
$\sigma^x$ state which differs from $\sigma$ by flipping spin at site
$x$. Therefore $\sigma^x(y)=(1-2\delta_{xy})\sigma(y)$.  

\begin{figure}
\onefigure[scale=0.4]{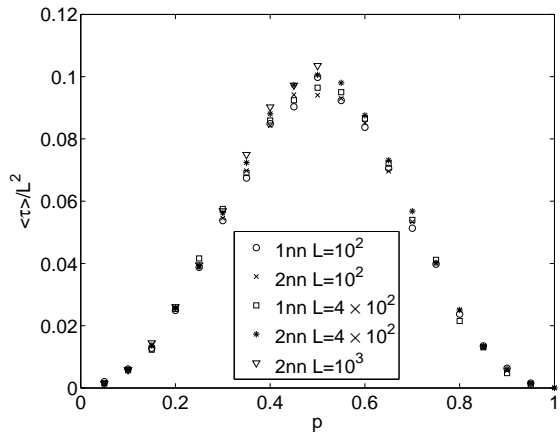}
\caption{The mean relaxation times from random initial state
consisting of $p$ up-spins for the modified (1nn) and original (2nn) outflow dynamics in one dimension for several lattice sizes $L$. In the modified version at most one spin is flipped in one step while in original formulation two can be flipped simultaneously. Therefore in case of modified version the time was rescaled by factor $\frac{1}{2}$. This is clearly visible the mean relaxation time scales with the lattice size as $\sim L^2$
analogously to the voter model \cite{L99,K92,MPR07}. The results
presented on the plot are averaged over $10^4$ samples.}
\label{fig:relaxtime}
\end{figure}

We introduce here slight modifications with respect to original outflow rule:
choose pair of neighbors and if they both are in the same state, then 
adjust one (instead of two) of its neighbors (chosen randomly on left or right with
equal probability $1/2$) to the common state. Because this way at most one spin is flipped in one step while in original formulation two can be flipped simultaneously, the time
must be rescaled by factor $\frac{1}{2}$. We measure the time so that the speed of
all processes remains constant when $L\to\infty$, so normally one update takes time $\frac{1}{L}$. Here, instead, we consider also the factor $\frac{1}{2}$, so single update takes time $\Delta t = \frac{1}{2N}$. Our modification eliminates some correlations due to
simultaneous flip of spins at distance $3$. However, if we look at
later stages of the evolution , where typically the domains are larger
than $2$, simultaneous flips occurs very rarely. Therefore, we do not expect any substantial difference. Indeed, computer simulations confirm our expectations - only time has to be rescaled (see Fig. \ref{fig:relaxtime}).

On the other hand, the modification simplifies the analytical
treatment. Indeed, the update rule can be equivalently formulated as
follows:
\begin{quote}
Choose randomly a spin $x$ and side $s$ ($s=1$ for right, $s=-1$ for left.) The
updated state is $\sigma(x;t+\Delta t)=\sigma(x+s;t)$ if
$\sigma(x+s;t)=\sigma(x+2s;t)$, otherwise $\sigma(x;t+\Delta t)=\sigma(x;t)$.
\end{quote}

Within such a formulation the probability that in one update the flip $\sigma\to\sigma^x$
occurs:

\begin{eqnarray}
W(\sigma \to \sigma^x) &=& 
\frac{1}{8N} \Big[ \sigma(x+2)\sigma(x+1)\nonumber\\
&+&\sigma(x-1)\sigma(x-2) - \nonumber\\
&-& \sigma(x) \Big( \sigma(x+2)+\sigma(x+1)+\sigma(x-1)+ \nonumber\\
&+& \sigma(x-2)  \Big)+2  \Big]
\label{eq:transprob}
\end{eqnarray}

These flip probabilities are then inserted into master equation:

\begin{equation}
P(\sigma;t+\Delta t)=\sum_{\sigma'}W(\sigma'\to\sigma)P(\sigma';t)
\end{equation}

Now, we make the limit $L\to\infty$ which also implies the continuous
time limit, as $\Delta t\to 0$. We also note that most of the
transition probabilities $W(\sigma'\to\sigma)$ are zero, since only
one spin flip is allowed in one step. Finally we end with
\begin{equation}
\frac{\mathrm{d}}{\mathrm{d}t}P(\sigma;t)=
\sum_x\Big[w(\sigma^x\to\sigma)P(\sigma^x;t)-w(\sigma\to\sigma^x)P(\sigma;t)]
\end{equation}
where the transition rates are trivially related to transition
probabilities (\ref{eq:transprob}),
$w(\sigma^x\to\sigma)=2NW(\sigma^x\to\sigma)$. (The sum is now over
infinite set of sites.)
For completeness we repeat the formula for transition rates:
\begin{eqnarray}
w(\sigma \to \sigma^x) & = & \frac{1}{4}\Big[
\sigma(x+2)\sigma(x+1)+\sigma(x-1)\sigma(x-2) \nonumber \\
&-&\sigma(x)\Big(\sigma(x+2)+\sigma(x+1)+\sigma(x-1) \nonumber \\
&+&\sigma(x-2)\Big)+2\Big]\;.
\label{eq:transrates}
\end{eqnarray}

It is hopeless to solve the master equation as it is. Instead, we write
evolution equations for some correlation functions derived from it. We
define:
\begin{eqnarray}
C_0(t)& = & \langle\sigma(y)\rangle\equiv\sum_{\sigma}\sigma(y)P(\sigma;t) \nonumber\\
C_1(n;t) & = & \langle\sigma(y)\sigma(y+n)\rangle \nonumber\\
C_2(n,m;t) & = & \langle\sigma(y-n)\sigma(y)\sigma(y+m)\rangle \nonumber\\
C_3(n,m,l;t) & = & \langle\sigma(y-n)\sigma(y)\sigma(y+m)\sigma(y+m+l)\rangle \nonumber\\
\vdots
\label{corr}
\end{eqnarray}

Only two equations are relevant for us. The first is:
\begin{equation}
\frac{\mathrm{d}}{\mathrm{d}t}C_0(t)=
-C_2(1,1;t)+C_0(t)
\label{eq:forRzero}
\end{equation}
and the second:
\begin{equation}
\frac{\mathrm{d}}{\mathrm{d}t}C_1(1;t)=
-C_3(1,1,1;t)-C_1(1;t)+C_1(3;t)+1
\label{eq:forRone}
\end{equation}
These two become closed set of equations, if we apply the
approximations described in the next section. Before going to it, it
is perhaps instructive to show the intermediate results which lead to
equations (\ref{eq:forRzero}), (\ref{eq:forRone}), 
and analogically to others, for more complicated
correlation functions.   

So, for example, for the lowest correlation function - the average of
one spin - we have
\begin{equation}
\frac{\mathrm{d}}{\mathrm{d}t}\langle\sigma(y)\rangle
=-2\langle w(\sigma\to\sigma^y)\sigma(y)\rangle
\end{equation}

and for the next one in the level of complexity 
\begin{eqnarray}
\frac{\mathrm{d}}{\mathrm{d}t}\langle\sigma(y)\sigma(y+1)\rangle 
 =  -2\langle w(\sigma\to\sigma^y)\sigma(y)\sigma(y+1)\rangle \nonumber\\
 -  2\langle w(\sigma\to\sigma^{y+1})\sigma(y)\sigma(y+1)\rangle\; .
\end{eqnarray}
The pattern is transparent. When computing the correlation function of
spins at sites $x_1$, $x_2$, $x_3$, \ldots, on the RHS we have sum of
terms, in which we average the product of spins at sites $x_1$, $x_2$,
$x_3$, \ldots with transition rate (which is constructed from the
spin configuration according to (\ref{eq:transrates})) for flip at positions $x_1$, $x_2$,
$x_3$, \ldots. As a formula, this sentence means
\begin{eqnarray}
\frac{\mathrm{d}}{\mathrm{d}t}\langle\prod_i \sigma(x_i)\rangle=-2\sum_j\langle w(\sigma\to\sigma^{x_j})\prod_i \sigma(x_i)\rangle\;.
\end{eqnarray}

\subsection{Kirkwood approximation}

Now we discuss the approximations used for solving
Eqs. (\ref{eq:forRzero}) and (\ref{eq:forRone}). 

The first one is the usual Kirkwood approximation, or decoupling,
which is used in various contexts and accordingly it assumes different
names. For example in the classical quantum many-body theory of
electrons and phonons in solids, it is nothing else than the
Hatree-Fock approximation (but contrary to this theory, which may be
improved systematically using diagrammatic techniques, here the
systematic expansions are not developed). We use the name Kirkwood
approximation, following the work \cite{mob_red_03}.

In our case, the Kirkwood approximation amounts to 
\begin{equation}
C_3(1,1,1;t)\simeq \big(C_1(1;t)\big)^2
\label{eq:Kirkwoodone}
\end{equation}
in Eq. (\ref{eq:forRone}) and 
\begin{equation}
C_2(1,1;t)\simeq C_1(1;t)C_0(t)
\label{eq:Kirkwoodzero}
\end{equation}
in Eq. (\ref{eq:forRzero}). While the latter assumption 
(\ref{eq:Kirkwoodzero})
enables us to relate the equation (\ref{eq:forRzero}) directly to
(\ref{eq:forRone}) and therefore to solve it as soon as we have the
solution of (\ref{eq:forRone}), the approximation
(\ref{eq:Kirkwoodone}) does not yet make of (\ref{eq:forRone}) a
closed equation. The point is that there is also the correlation at
distance $3$, the function $C_1(3;t)$. So, we make also an additional
approximation, which is also made in  \cite{mob_red_03}. We suppose
that $C_1(n;t)$ only weakly depends on distance $n$, or else, that the
decay of the correlations is relatively slow. If the spins are
correlated to certain extent on distance $1$ (the neighbors) , they
are correlated to essentially the same extent also on distance $3$
(next-next neighbors). This is also justified if the domains are
large enough, i. e. at later stages of the evolution. So, we assume
\begin{equation}
C_1(3;t)\simeq C_1(1;t)\;.
\label{eq:slowdecay}
\end{equation}

\begin{figure}
\onefigure[scale=0.4]{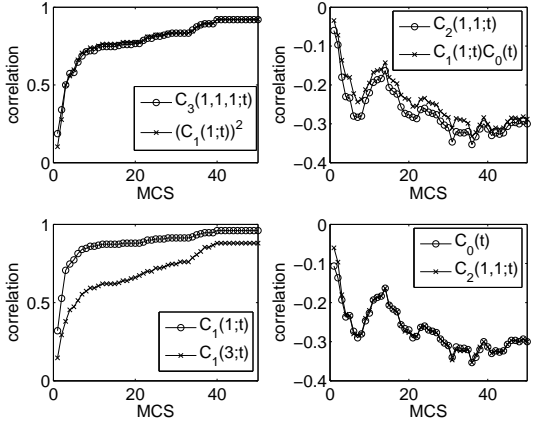}
\caption{Sample (not averaged, one simulation only) time evolution of several correlation functions given by eq. (\ref{corr}) for random initial conditions with $p$ up-spins. Kirkwood approximation given equations(\ref{eq:Kirkwoodone}), (\ref{eq:Kirkwoodzero}) and assumption (\ref{eq:slowdecay}) are valid for later stages, although the assumption (\ref{eq:Kirkwoodone}) agrees perfectly with simulation results from very beginning (left upper panel).}
\label{fig:corr1_r}
\end{figure}

In the figure \ref{fig:corr1_r} we present a sample (not averaged) time evolution of several correlation functions. Indeed, our assumptions can be justified at later stages of the evolution, although the assumption (\ref{eq:Kirkwoodone}) agrees perfectly with simulation results from very beginning. The second given by (\ref{eq:Kirkwoodzero}) agrees with simulations also quite well. Only the assumption (\ref{eq:slowdecay}) that the decay of the correlations is relatively slow is valid only at later stages of the evolution. 

To sum it up, the approximations (\ref{eq:Kirkwoodone}),
(\ref{eq:Kirkwoodzero}), and (\ref{eq:slowdecay}) say that
approximately
\begin{eqnarray}
C_0(t) & \simeq & \psi(t) \nonumber\\
C_1(n;t)& \simeq &\phi(t)
\end{eqnarray}
where $\psi(t)$ and $\phi(t)$ satisfy  the equations (dot denotes the time-derivative
\begin{eqnarray}
\dot{\psi} & = & (1-\phi)\psi \nonumber\\
\dot{\phi} & = & 1-\phi^2\; .
\label{eq:phipsiset}
\end{eqnarray}
The solution is straightforward. We assume initial conditions
$\phi(0)=m_1$ and $\psi(0)=m_0$. First we solve the second equation
from the set (\ref{eq:phipsiset}). This gives
\begin{equation}
\phi(t)=\frac{\sinh t+ m_1 \cosh t}{\cosh t+ m_1\sinh t}
\end{equation}
and inserting that into the first of the set  (\ref{eq:phipsiset}) we
have
\begin{equation}
\psi(t)=\frac{2m_0}{1+m_1+(1-m_1)\,\mathrm{e}^{-2t}}\;.
\end{equation}
The most important result is the asymptotics
\begin{equation}
\psi(\infty)=\frac{2m_0}{1+m_1}\;.
\label{eq:psiinfty}
\end{equation}

How to interpret this finding? The average $C_0(t)$ is the average
magnetization. In other terms, it determines the probability that a
randomly chosen spin will have state $+1$ at time $t$. This
probability is $p_+(t)=(C_0(t)+1)/2$. Therefore, $m_0=C_0(0)$ is the
initial magnetization. When we go to the limit $t\to\infty$, we know
that ultimately the homogeneous state is reached. The asymptotic
magnetization $C_0(\infty)$ therefore says what is the probability
that the final state will have all spins $+1$. It is
$(C_0(\infty)+1)/2$. So, (\ref{eq:psiinfty}) means that
\begin{equation}
C_0(\infty)\simeq\frac{2C_0(0)}{1+C_1(1;0)}\;.
\label{eq:finc0}
\end{equation}
If the initial state is completely uncorrelated, i. e. we set the
spins at random, with the only condition that average magnetization
is $m_0$, we have $C_1(1;0)=m_0^2$ and
\begin{equation}
C_0(\infty)\simeq\frac{2m_0}{1+m_0^2}\;.
\end{equation}
Finally, we express this result in terms of the probability
$p=(C_0(0)+1)/2$ 
to have a
randomly chosen spin spin in state $+1$ at the beginning and the
probability $P_+=(C_0(\infty)+1)/2$ that all spins are in state $+1$ at the end. We have
\begin{equation}
P_+\simeq\frac{p^2}{2p^2-2p+1}\;.
\label{exitp}
\end{equation}

Computer simulations for random initial conditions, in which
assumption $C_1(1;0)=m_0^2$ can be done shows perfect agreement with
analytical formula (\ref{exitp}). In the next section we show how
results will change in case of correlated initial conditions. 

\section{Correlated initial conditions}
Here we consider two examples of correlated initial conditions with fraction $p$ of up-spins:
\begin{enumerate}
\item
Ordered initial state that consists of two clusters: $pL$-length of up-spins and $(1-p)L$-length of  
down-spins, for example in case of $L=10$:
\begin{eqnarray}
p=0.5 &:&  \uparrow \uparrow \uparrow \uparrow \uparrow 
\downarrow \downarrow \downarrow \downarrow \downarrow \nonumber\\
p=0.4 &:&  \uparrow \uparrow \uparrow \uparrow \downarrow 
\downarrow \downarrow \downarrow \downarrow \downarrow \nonumber\\
p=0.3 &:&  \uparrow \uparrow \uparrow \downarrow \downarrow 
\downarrow \downarrow \downarrow \downarrow \downarrow \nonumber\\
\ldots
\end{eqnarray}
\item
Correlated, completely homogeneous, initial state, i.e. $S_{n/p}=1$ for $n=0,1,2,3,...$,
for example in case of $L=8$:
\begin{eqnarray}
p=0.5 &:&  \uparrow \downarrow \uparrow \downarrow \uparrow \downarrow \uparrow \downarrow \nonumber\\
p=0.25 &:&  \downarrow \downarrow \downarrow \uparrow \downarrow \downarrow \downarrow \uparrow\\
\ldots
\end{eqnarray}
\end{enumerate}

\begin{figure}
\onefigure[scale=0.4]{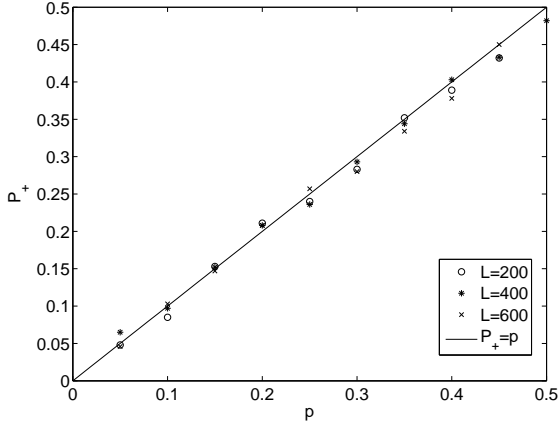}
\caption{Exit probability $P_+$ from ordered initial state
consisting of $p$ up-spins for the outflow dynamics in one dimension
for several lattice sizes $L$. Initial state consists of two clusters:
$pL$-length of up-spins and $(1-p)L$-length of  down-spins. Results
for original and modified rules are the same. The dependence between
initial ratio of up-spins $p$ and exit probability is given by the
simplest linear function $P_+=p$ as in the case of the voter
model. Analytical result in this case can be obtained from the
equation (\ref{eq:finc0}). In this case $C_1(1;0) = 1-\frac{1}{L}
\approx 1$ and we obtain $C_0(\infty)=c_0(0)=m_0$, i.e. $P_+=p$, which
perfectly agree with simulation results. Results are averaged over
$10^3$ samples.} 
\label{fig:exitprob_order}
\end{figure}

In both cases it is easy to calculate exactly correlation function $C_1(1;0)$. In the first case of ordered initial conditions we obtain:
\begin{equation}
C_1(1;0) = 1-\frac{1}{L} \approx 1
\end{equation} 
Thus, from equation (\ref{eq:finc0}):
\begin{equation}
C_0(\infty)\simeq\frac{2C_0(0)}{1+C_1(1;0)}=\frac{2m_0}{1+1}=m_0
\rightarrow P_+=p. 
\label{eq:exitprob_order}
\end{equation}
Computer simulations shows that indeed for such a initial conditions
$P_+=p$ (see Fig. \ref{fig:exitprob_order}). 

\begin{figure}
\onefigure[scale=0.4]{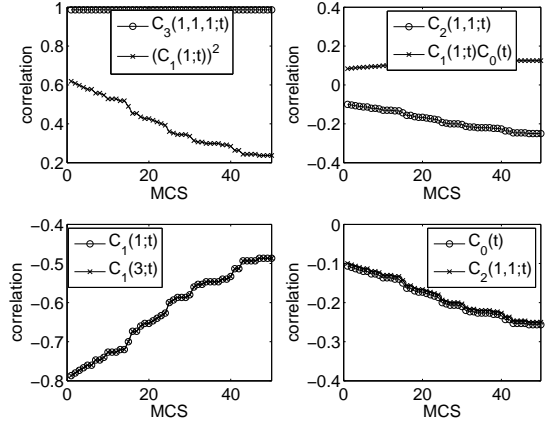}
\caption{Sample (not averaged, one simulation only) time evolution of
  several correlation functions given by eq. (\ref{corr}) for random
  ordered initial conditions with $p$ up-spins. Initial state consists
  of two clusters: $pL$-length of up-spins and $(1-p)L$-length of
  down-spins. Kirkwood approximation given
  equations(\ref{eq:Kirkwoodone}) and (\ref{eq:Kirkwoodzero}) are not
  valid.} 
\label{fig:corr_order}
\end{figure}

As we see Kirkwood approximation surprisingly gives correct results
also in this case. However, if we look at figure \ref{fig:corr_order}
we see that Kirkwood approximation given equations by
(\ref{eq:Kirkwoodone}) and (\ref{eq:Kirkwoodzero}) cannot be justified
by computer simulations.  

\begin{figure}
\onefigure[scale=0.4]{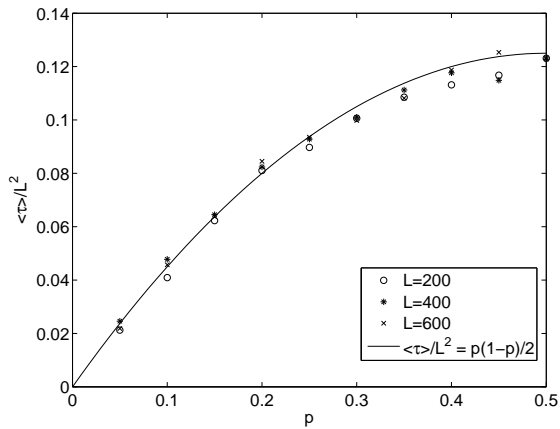}
\caption{The mean relaxation times for the outflow dynamics in one
  dimension for several lattice sizes $L$. Initial state consists of
  two clusters: $pL$-length of up-spins and $(1-p)L$-length of
  down-spins. Results for original and modified rules are the
  same. This is clearly visible that in case of such an ordered
  initial state the dependence between initial ratio of up-spins $p$
  and the mean relaxation time $\langle\tau\rangle$ is given by simple parabola
  not a bell-shaped curve ( like in case of the random initial
  conditions). However, still the mean relaxation time scales with the
  lattice size $ \sim L^2$. The results presented on the plot are
  averaged over $10^4$ samples. } 
\label{fig:meantime_order}
\end{figure}

We have checked also the mean relaxation time in case of ordered
initial conditions (Fig. \ref{fig:meantime_order}). It occurs that
analogously like for random initial conditions the mean relaxation
time scales with the system size as $\langle\tau\rangle \sim L^2$ (see
Figs. \ref{fig:relaxtime} and \ref{fig:meantime_order}). The same
scaling has been found in the voter model \cite{L99,K92,MPR07}. 
However, contrary to the random initial conditions for which
bell-shaped curve is observed, here the mean relaxation times is well
described by simple parabola: 
\begin{equation}
\frac{\langle\tau\rangle}{L^2} = \frac{1}{2}p(1-p).
\end{equation}

For the second correlated initial conditions, which are completely
homogeneous we observed in computer simulations that exit probability
is step like, i.e.  
\begin{eqnarray}
P_+=0 & \mbox{ for $p<0.5$} \nonumber\\
P_+=1 & \mbox{ for $p>0.5$} \nonumber\\
\mbox{antiferromagnetic state}  & \mbox{ for $p=0.5$} 
\end{eqnarray}

In this case two-spins correlation function can be also calculated
easily. For $p=\frac{1}{n}<0.5, n \in N$ we obtain: 
\begin{equation}
C_1(1;0) = p(1 \times (\frac{1}{p}-2) + (-1) \times 2) = p(\frac{1}{p}-4) = 1-4p.
\end{equation} 
Thus, from equation (\ref{eq:finc0}):
\begin{equation}
C_0(\infty)\simeq\frac{2C_0(0)}{1+C_1(1;0)}=\frac{4p-2}{2-4p}=-1
\rightarrow P_+=0, 
\end{equation}
which again agrees perfectly with computer simulations, although
Kirkwood approximation cannot be easily justified. 

\begin{figure}
\onefigure[scale=0.4]{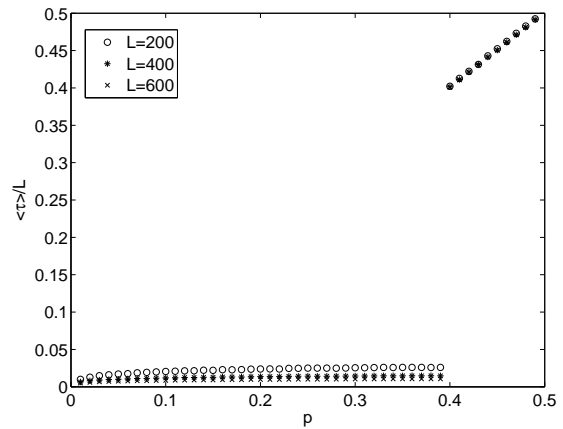}
\caption{The mean relaxation times for the outflow dynamics in one
  dimension for several lattice sizes $L$. Initial state consists of
  $p$ up-spins and it is completely homogeneous, i.e. $S_{n/p}=1$ for
  $n=0,1,2,3,...$ (e.q. for $p=1/4$ the initial state is  
$\uparrow \downarrow \downarrow \downarrow \uparrow \downarrow
  \downarrow \downarrow \uparrow \downarrow \downarrow
  \downarrow$). Results for original and modified rules are the
  same. In such a case scaling $\langle\tau\rangle \sim L^2$ is not
  valid anymore 
  and for $p<0.4 (p>0.4)$ the relaxation time does not depend on the
  lattice size. However, for $p \in (0.4,0.6)$ the mean relaxation
  time scales as $\langle\tau\rangle \sim L$ and depends linearly on
  the initial 
  ratio of up-spins $p$. The results presented on the plot are
  averaged over $10^4$ samples. } 
\label{fig:relaxtime_homo}
\end{figure}

Very interesting results is obtained for homogeneous initial condition
if we measure the mean relaxation time (see
Fig. \ref{fig:relaxtime_homo}). Computer simulation shows that for
$p<0.4$ (and $p>0.6$, respectively) the mean relaxation time
$\langle\tau\rangle$ 
does not depend on the system size $L$ and for $p \in (0.4,0.6)$
depends linearly on  system size,
i.e. $\langle\tau\rangle \sim L$.  

\section{Summary}
We introduced modified version of one-dimensional outflow dynamics in which we 
choose pair of neighbours and if they both are in the same state, then 
adjust {\it one} (in original version both) of its neighbours (chosen
randomly on left or right with 
equal probability $1/2$) to the common state. We checked in computer
simulations that accordingly to our expectations results in the case
of modified rule are the same as in the case of original outflow
dynamics, only the time must be rescaled by factor $\frac{1}{2}$. 
Modified version simplified the analytical treatment and allowed to
derive the master equation. Following the method proposed in
\cite{mob_red_03} we wrote evolution equations for some correlation
functions and used the Kirkwood approximation. This approach allowed
us to derive the analytical formula for final magnetization
(\ref{eq:finc0}). 
In fact, just before finishing this paper the same result was published by
Lambiotte and Redner as a special case in the work \cite{lam_red_07} 
where a model interpolating the voter,
Majority-rule (or Sznajd) and so-called vacillating voter dynamics was
investigated, using also the Kirkwood approximation.

In the case of random initial conditions Kirkwood
approximation can be justified looking at time evolution of simulated
correlation functions. In this case our analytical results can be
simplified to eq. (\ref{exitp}) and agrees perfectly with simulations
on contrary to earlier approaches \cite{g05,GSH2006}. We have checked
also how the Kirkwood approximation works in the case of two types of
correlated initial conditions. Although in both cases the Kirkwood
approximation cannot be easily justified, surprisingly we obtained
perfect agreement with computer simulations.  

\acknowledgments
This work was supported by the M\v SMT of the Czech Republic, grant no. 
1P04OCP10.001, and by the Research Program CTS MSM 0021620845. 

Katarzyna Sznajd-Weron gratefully acknowledges the financial
support in period 2007-2009 of the Polish Ministery of Science and
Higher Education 
through the scientific grant no. N N202 0194 33

\end{document}